\documentclass[aps,twocolumn,showpacs,preprintnumbers,nofootinbib,prl,superscriptaddress,groupedaddress,10pt]{revtex4-1}

\makeatletter
\def\l@subsubsection#1#2{}
\def\l@subsubsubsection#1#2{}
\makeatother

\usepackage{graphicx,amssymb,amsmath,amsthm,amsfonts,epsfig,epsf}
\usepackage[usenames,dvipsnames,svgnames,table]{xcolor}
\usepackage{epstopdf}
\definecolor{darkred}{rgb}{0.5,0,0}
\usepackage{aas_macros}
\usepackage{bm}
\usepackage{dcolumn}
\usepackage[latin1]{inputenc}
\usepackage{latexsym}
\usepackage{rotating}
\usepackage{longtable}

\usepackage{enumerate}
\usepackage{tensor,multirow}
\usepackage{url}
\usepackage[linktocpage]{hyperref}
\usepackage{float}
\usepackage{hyperref}

\def\be{\begin{equation}}
\def\ee{\end{equation}}
\newcommand{\beq}{\begin{eqnarray}}
\newcommand{\eeq}{\end{eqnarray}}

\def\ba{\begin{align}}
\def\ea{\end{align}}

\newcommand{\red}[1]{{\textcolor{black}{#1}}}

\newcommand{\TPsi}{\tilde{\Psi}}

\begin{document}
\title{Characterization of echoes:\\
A Dyson-series representation of individual pulses}

\author{
Miguel R. Correia$^{1}$,
Vitor Cardoso$^{1,2}$
}
\affiliation{${^1}$ CENTRA, Departamento de F\'{\i}sica, Instituto Superior T\'ecnico -- IST, Universidade de Lisboa -- UL,
Avenida Rovisco Pais 1, 1049 Lisboa, Portugal}
\affiliation{${^2}$ Perimeter Institute for Theoretical Physics, 31 Caroline Street North Waterloo, Ontario N2L 2Y5, Canada}
%
\begin{abstract}
The ability to detect and scrutinize gravitational waves from the merger and coalescence of compact binaries
opens up the possibility to perform tests of fundamental physics. One such test concerns the dark, nature of compact objects:
are they really black holes? It was recently pointed out that the absence of horizons -- while keeping the external geometry very close to that of General Relativity -- would manifest itself in a series of echoes in gravitational wave signals. 
The observation of echoes by LIGO/Virgo or upcoming facilities would likely inform us on quantum gravity effects or unseen types of matter. Detection of such signals is in principle feasible with relatively simple tools, but would benefit enormously from accurate templates. Here we analytically individualize each echo waveform and show that it can be written as a Dyson series, for arbitrary effective potential and boundary conditions. We further apply the formalism to explicitly determine the echoes of a simple toy model: the Dirac delta potential. Our results allow to read off a few known features of echoes and may find application in the modelling for data analysis. 
\end{abstract}

\maketitle

\tableofcontents

\section{Introduction}
Black holes (BHs) are one of the most intriguing solutions of Einstein field equations.
They play a key role in the understanding of the mathematical content of the field equations,
but are also a fundamental player in astrophysics and -- within the gauge-gravity duality context -- in high energy physics.
In General Relativity (GR), BHs are very simple and characterized by a one-way membrane called the event horizon. The event horizon and associated boundary conditions is the chief responsible for a number of properties of BHs, including their simplicity~\cite{Cardoso:2016ryw}. In addition, the ``one-way'' property of horizons causally disconnects the BH interior from its exterior, shielding
outside observers from unknown quantum effects generated close to singularities or those presumably going on at the Cauchy horizon
of spinning or charged BHs. Nevertheless, such extraordinary properties naturally raise the question of whether horizons do indeed {\it always} form, and where is this dynamical mechanism hiding in the field equations~\cite{Cardoso:2017njb,Berthiere:2017tms}.

Other questions relate to the nature of compact objects and to tests of gravity in regions of where spacetime becomes so warped
as to potentially form horizons and hide singularities, regions with huge intrinsic redshift.
There are numerous tests of gravity in the ``weak field'' regime, and GR seems to successfully describe these regions up to the currently accessible precision. What about strongly warped spacetime regions? What evidence do we have that they exist, how is it quantified, and how can we quantify the evidence for horizons?

These questions all meet at horizons. Fortunately, the access to such fundamental questions has recently been granted through the historical detection of gravitational waves (GWs) by aLIGO~\cite{Abbott:2016blz}.
Compact binaries are the preferred sources for GW detectors (and in fact were the source of the first detected events~\cite{Abbott:2016blz,Abbott:2016nmj}). 
The GW signal from compact binaries is naturally divided
in three stages, corresponding to the different cycles in the evolution driven by GW emission~\cite{Buonanno:2006ui,Berti:2007fi,Sperhake:2011xk}: the inspiral stage, corresponding to large separations and well approximated by post-Newtonian theory; the merger phase when the two objects coalesce and which can only be described accurately through numerical simulations; and finally, the ringdown phase when the merger end-product relaxes to a stationary, equilibrium solution of the field equations~\cite{Sperhake:2011xk,Berti:2009kk,Blanchet:2013haa}.

\subsection{Quantifying the evidence for horizons with precision GW-physics:\\
Multipoles, tidal heating and tidal Love numbers}
All three stages provide independent, unique tests of gravity and of compact GW sources. Due to their very nature, it is impossible to 
{\it prove} the existence of event horizons~\cite{Abramowicz:2002vt,Cardoso:2017njb}. However, the evidence for horizons and BHs may be significantly tightened using GW observations:

During the inspiral, three different features will play a role, all of which can be used to test the BH-nature of the objects. 
Firstly, horizonless objects are bound to possess a \emph{different multipolar structure from Kerr BHs,} impacting the GW phase,
and allowing for constraints on possible deviations from the Kerr geometry~\cite{Krishnendu:2017shb}.
Secondly, the presence of horizons translates into BHs absorbing radiation in a very characteristic way.
Horizons absorb incoming high frequency radiation, and serve as sinks or amplifiers for low-frequency radiation able to tunnel in.
However, horizonless objects are expected not to have significant \emph{tidal heating}. Thus, a ``null-hypothesis'' test consists on using the phase of GWs to measure absorption or amplification at the surface of the objects~\cite{Maselli:2017cmm}. LISA-type GW detectors~\cite{Audley:2017drz} will place stringent tests on this property, potentially reaching Planck scales near the horizon and beyond~\cite{Maselli:2017cmm}.
Finally, non- or slow-spinning BHs were shown to have zero \emph{tidal Love numbers}. In a binary, the gravitational pull of one object deforms its companion, inducing a quadrupole moment proportional to the tidal field.
The tidal deformability is encoded in the Love numbers, and the consequent modification of the dynamics can be directly translated into GW phase evolution at higher-order in the post-Newtonian expansion~\cite{Flanagan:2007ix}. 
It turns out that the tidal Love numbers of a BH are zero~\cite{Binnington:2009bb,Damour:2009vw,Fang:2005qq,Gurlebeck:2015xpa,Poisson:2014gka,Pani:2015hfa}, allowing again for null tests. By devising these tests, existing and upcoming detectors can rule out or strongly constraint boson stars~\cite{Wade:2013hoa,Cardoso:2017cfl,Maselli:2017cmm,Sennett:2017etc} or even generic ultracompact horizonless objects~\cite{Cardoso:2017cfl,Maselli:2017cmm}.

\subsection{Quantifying the evidence for horizons with smoking guns: GW echoes}
The GW signal overall depends on the entire spacetime structure. The late-time behavior depends, mostly,
on the strong-field region and on the boundary conditions set there. Generically, a BH-type ringdown is excited when the spacetime is disturbed at the photosphere. We call this the photosphere ringdown, which has a timescale of order 
$\sim 5GM/c^3$~\cite{Cardoso:2017njb} (we will use geometric units, $G=c=1$, from now on). If light or GWs hit the surface of the putative horizonless objects on timescales of this order or smaller, then the ringdown of the spacetime is significantly different from that of a BH. Thus, tests of GR or of the nature of the object can be performed using the standard ringdown tools~\cite{Berti:2005ys,Berti:2016lat,Cardoso:2016ryw}.

These techniques and tools require very sensitive detectors and hinge on the ability to do {\it precision physics}.
It turns out that, should horizons {\it not} exist, there should be clear signatures of this absence in GW signals.
If the outside spacetime is vacuum GR all the way down to a region very close to the horizon, in a way that GWs take a time $\gg 5M$ to hit the surface, then we call these Clean Photosphere Objects (ClePhOs). Because ClePhOs have photospheres
which are vacuum in their vicinity, the photosphere ringdown stage is therefore indistinguishable from that of a BH~\cite{Cardoso:2016rao,Cardoso:2017njb}. However, on longer timescales the GWs excited at the photosphere have had time to hit the surface and bounce back to the observer, with a fraction of it trapped between the object and the photosphere:
the late-time response of a ClePhO is a series of GW echoes~\cite{Cardoso:2016rao,Cardoso:2017cqb,Cardoso:2017njb,Cardoso:2016oxy,Price:2017cjr,Volkel:2017ofl,Volkel:2017kfj}. From this discussion it is clear that echoes should also appear whenever there is structure in the near horizon region, either due to the compact object or due to propagation effects~\cite{Zhang:2017jze}. 

\subsection{The morphology of echoes and purpose of this work}

The non-detection of echoes can place interesting bounds on the surface of ClePhOs and rule out certain models; most importantly, it would {\it quantify} in an independent way the evidence for BHs in our universe.
Given the tremendous significance of a possible detection of echoes in a GW signal, it is worth
devising strategies to detect them in actual stretches of data. Early studies, based on simple templates,
have claimed an important statistical evidence for the presence of echoes in the first detections~\cite{Abedi:2016hgu,Abedi:2017isz}.
At this point the evidence is not of enough statistical significance to warrant claims of a detection~\cite{Ashton:2016xff,Westerweck:2017hus}. To improve the significance and to understand the underlying physics, a better description of GW echoes is necessary.

An important step in this direction was given recently by Mark and collaborators~\cite{Mark:2017dnq} 
(see also Refs.~\cite{Nakano:2017fvh,Bueno:2017hyj}). These authors have shown how the response of a ClePho can be expressed in terms of the response of a BH, convoluted with an appropriate function which takes into account the different boundary conditions. Doing this, they were able to identify the contribution from isolated echoes and proposed templates for non-spinning BHs which had a very good match with numerical waveforms. Subsequent work has investigated phenomenological waveforms for detection of echoes~\cite{Maselli:2017tfq}. Clearly, the result of these incursions is that a lot more remains to understand. Some of the open issues include,
\begin{itemize}

\item The very late time behavior of the GW response of ClePhOs is described by the quasinormal modes (QNMs) of the system.
The QNMs arise as poles of the relevant Green's function. It is a puzzle that the photosphere modes have no special spectral role, and yet they dominate the response at early times. 

\item In addition, although the late-time response of ClePhOs is presumably governed by its QNMs, no numerical evidence exists of this fact.

\item The overall amplitude of sucessive echoes decreases, at least if one is looking for consecutive echoes generated shortly after merger. What type of decay is this, is it polynomial, exponential? Can we characterize the evolution of echoes in a more precise manner? 

\item The delay between different echoes is a key quantity in any detection strategy. Is the delay really constant or does it evolve in time, and how~\cite{Wang:2018mlp}?

\item In a related vein, a generic widening of the pulses, in the time-domain, was observed as time goes by. This is physically intuitive: the pulses are semi-trapped within a cavity that let's high frequency waves pass. At late times only low-frequency,
resonant modes remain. Hence the pulse is becoming more monochromatic. This is a generic but not yet quantified result.

\item Consecutive echoes may be in phase or out of phase, depending on the particular boundary conditions imposed on us by the physical model. 

\end{itemize}
\par
The purpose of this work is to start the discussion on some of these issues. Here we provide an exact expression for each individual echo amplitude, in the form of a Dyson series, that accounts for most of the above points. This formula also works retrospectively: if one echo is measured, then the surface reflectivity becomes fully specified (assuming the potential is known). 
\par
The first section consists of the perturbative approach to wave scattering in which: i) the most general appropriate boundary conditions are considered; ii) the Lippmann-Schwinger equation is obtained; iii) the resulting Dyson is resummed into a sum of \emph{echoes}; iv) the time-dependent solution is obtained through Laplace inversion.
\par
In the second section, we apply this apparatus to a Dirac delta type of potential. The Dyson series does not need to be truncated since the explicit solution is completely attainable. Here we compute the echo frequency amplitude for arbitrary surface reflectivity, and the echo waveform for a set of common boundary conditions.  

\section{Waves in open systems}

A linear perturbation $\Psi(t,x)$ in a system with potential $V(x)$, supported on $\mathbb{R}$, obeys the wave equation  
\begin{equation} \label{eq:1}
-{\partial^2 \Psi \over \partial t^2} + {\partial^2 \Psi \over \partial x^2} - V(x) \Psi = 0 \, .
\end{equation}
\par
\subsection{Boundary Conditions}
We're interested in \emph{open systems}, in which waves are only partially bounded and can escape to infinity, in at least of one of the sides.
We'll choose $+ \infty$ to be the open end. Then, 
\be \label{eq:bc1}
\Psi(t,x \to  \infty) \propto e^{ i \omega ( x- t)} \, ,
\ee 
The other boundary condition (BC) can correspond to a reflection at some point $x = - L$. When the potential vanishes it can be written as
\be
\label{eq:bc}
\Psi(t,x \sim - L) \propto e^{-i \omega(x+t)} + R(\omega) \,  e^{i \omega(x-t)} \, ,
\ee
where the first term on the rhs corresponds to a wave travelling to the left, out of the system, and the second term is nothing but the reflected wave, travelling to the right, thus, we can identifty $R(\omega)$ as the \emph{reflectivity} associated with the BC at $x = -L$.

If we do not wish for external influence on the system, the reflectivity should obey
\be
\label{eq:R}
|R(\omega)| \leq 1 \, ,
\ee
otherwise, the reflected wave has a larger amplitude than the outgoing wave, i.e. there's an external input at the left.
The condition above is however violated for some well-known systems, such as Kerr BHs, which are known to display superradiance~\cite{Brito:2015oca}.

The reflection coefficient $R(\omega)$ is completely specified by the BC at $x = -L$.
We can point out three familiar cases. For a purely outgoing wave, to the left, we simply have $R(\omega) = 0$. For a Dirichlet BC, imposing $\Psi(t,-L) = 0$ on \eqref{eq:bc}, we have $R(\omega) = - e^{i \omega 2 L}$, whereas for a Neumann BC ($\partial_x \Psi(t,-L) = 0$) we get $R(\omega) = e^{i \omega 2 L}$. Both of the latter two are conservative, they satisfy $|R(\omega)| = 1$.

Conversely, $R(\omega)$ can be arbitrarily chosen and the BC at $x = -L$ becomes automatically imposed. For instance, dissipation can be introduced by generalizing the latter reflectivities to 
\be \label{eq:Rb}
R(\omega) = - r e^{i \omega 2 L}\,,
\ee
with $r \in [-1,1]$ and $|R(\omega)| = |r| \leq 1$. The BC at $x=-L$ turns out to be $\Psi(t,-L) \propto (1-r)e^{i \omega(L -t)}$ and $\partial_x \Psi(t,-L) \propto - i \omega (1+r)e^{i \omega(L -t)}$.
\subsection{The Dyson series solution of the Lippman-Schwinger equation}
To solve \eqref{eq:1} we employ the Laplace transform~\cite{Berti:2009kk}
\be
\tilde{\Psi}(\omega,x) = \int_{0}^\infty \Psi(t,x) e^{i \omega t} \, dt \, .
\ee
If $\Psi(t \to \infty) \sim e^{\alpha t}$ then $\tilde{\Psi}$ only converges for $\Im{\omega} > \alpha$. 
\par
The time-dependent solution is the inverse of this transform,
\be
\label{eq:inverse}
\Psi(t,x) = {1 \over 2 \pi} \int_{- \infty + i \beta}^{+ \infty + i \beta} \TPsi(\omega,x) e^{- i \omega t} \, d \omega\,,
\ee
where $\beta$ assumes any value $\beta > \alpha$ to ensure the integrand is always convergent along the path of integration.

With these definitions, Eq.~\eqref{eq:1} is reduced to the ODE
\begin{equation} 
{d^2 \TPsi \over dx^2} + \big( \omega^2 - V(x) \big) \TPsi = I(\omega,x) \,,\label{eq:10}
\end{equation}
with source term
\be
I(\omega, x) = i\omega \psi_0(x) - \dot{\psi}_0(x) \,,\label{eq:I}
\ee
where $\psi_0(x) = \Psi(0,x)$ and $\dot{\psi_0}(x) = \partial_t \Psi(0,x)$ are the initial conditions.

Now, instead of pursuing the usual Green's function approach, we use a perturbative framework. The ODE \eqref{eq:10}, and BCs, can be rewritten in the Lippmann-Schwinger integral form
\begin{equation} \label{eq:integral}
\TPsi(\omega,x) = \TPsi_0(\omega,x)+ \int_{- L}^{\infty} g(x,x') \, V(x') \TPsi(\omega,x') \, dx'\,,
\end{equation}
where 
\begin{equation} \label{eq:g}
g(x,x') = {e^{i \omega |x-x'|} + R(\omega) \, e^{i \omega (x+x')} \over 2 i \omega} \, ,
\end{equation}
is the Green's function of the free wave operator $d^2/d x^2 + \omega^2$ with BCs \eqref{eq:bc1} and \eqref{eq:bc}, and
\begin{equation} \label{eq:free}
\TPsi_0(\omega,x) = \int_{- L}^{\infty} \! g(x,x') \, I(\omega,x') \, dx'\,,
\end{equation}
is the free-wave amplitude.

The formal solution of Eq. \eqref{eq:integral} is the Dyson series
\begin{align} \label{eq:dyson}
\TPsi(\omega,x) = \sum_{k=1}^\infty \int_{-L}^{\infty}& \! g(x,x_1) \cdots g(x_{k-1},x_k) V(x_1) \nonumber \\
& \cdots V(x_{k-1}) I(\omega,x_k) dx_1 \cdots dx_k \, ,
\end{align}
which effectively works as an expansion in powers of $V/\omega^2$, so we expect it to converge rapidly for high frequencies. 

Note that if we were to expand each term of the series with explicit use of \eqref{eq:g} we'd get a panoply of powers of $R(\omega)$. Now, we may ask, is it possible to reorganize \eqref{eq:dyson} and express it as a series in powers of $R(\omega)$?
\subsection{Resummation of the Dyson series and echoing structure}
We start by separating the Green's function \eqref{eq:g} into $g = g_o + R\, g_r$, with
\be \label{eq:go}
g_o(x,x') = {e^{i \omega |x-x'|} \over 2 i \omega} \, ,
\ee
the open system Green's function, and
\be \label{eq:gr}
g_r(x,x') = {e^{i \omega (x+x')} \over 2 i \omega} \, ,
\ee
the ``reflection'' Green's function.

We can then write \eqref{eq:integral} as
\begin{align} \label{eq:step0}
\TPsi(\omega,x) = & \int_{- L}^{\infty} g_o(x,x') \, I(\omega,x') \, dx'  \nonumber \\ 
 + & R(\omega) \int_{- L}^{\infty} g_r(x,x')\, I(\omega,x') \, dx' \nonumber \\
+ &\int_{- L}^{\infty} g(x,x') \, V(x') \TPsi(\omega,x') \, dx' \, .
\end{align}
Now, in the same way as a Dyson series is obtained, we replace the $\TPsi(\omega,x')$ in the third integral with the entirety of the rhs of Eq. \eqref{eq:step0} evaluated at $x'$. Collecting powers of $R(\omega)$ yields
\begin{align} \label{eq:step1}
\TPsi= & \int \! g_o I + \int \!\!\int\! g_o V g_o    I  \nonumber \\ 
 + & R \, \bigg[ \int \! g_r I + \int \!\! \int\! (g_r V g_o + g_o V g_r)  I \bigg] \nonumber \\
+ & R^2 \!\int\!\! \int\! g_r V g_r I 
+ \int\!\! \int \! g \, V g \, V   \TPsi  \, ,
\end{align}
where, for better clarity, we chose not to write the functions' arguments. 
\par
If we repeat the process one more time, i.e. by replacing Eq. \eqref{eq:step0} with $\TPsi$ in the last integration in \eqref{eq:step1}, we get
\begin{align} \label{eq:step2}
\TPsi= & \int \! \!g_o I + \!\int \!\!\!\int\!\! g_o V g_o  I  + \!\int\!\!\!\int \!\!\!\int\!\! g_o V  g_o  V  g_o I \nonumber \\ 
 + & R \, \bigg[ \!\int \!\! g_r I + \!\int \!\!\! \int\! (g_r V g_o + g_o V g_r) I  \nonumber \\ 
 +& \!\int\!\!\!\int\! \!\!\int\! (g_o V g_o V g_r + g_o V g_r V g_o + g_r V g_o V g_o) I \bigg] \nonumber \\
+ & R^2 \, \bigg[ \!\int\!\!\! \int\! \!g_r  V g_r I +  \!\int\!\!\!\int \!\!\!\int\! (g_o V g_r V g_r + g_r V g_r V g_o + g_r V g_o V g_r) I \bigg] \nonumber \\ 
+ & R^3  \!\int\!\!\!\int \!\!\!\int\!\!  g_r V g_r V g_r I +  \!\int\!\!\!\int \!\!\!\int\!\! g \, V g \, V  g \, V \TPsi  \, ,
\end{align}
and a pattern starts to emerge. The first line does not contain any $g_r$, the factor of $R$ contains one $g_r$ arranged in all possible distinct ways with the $g_o$'s, the factor of $R^2$ contains two $g_r$'s also arranged in all possible ways, and so on and so forth. If we continue this process we end up with a geometric-like series in powers of $R$,

\begin{equation} \label{eq:tpsi}
\TPsi(\omega,x) = \TPsi_o(\omega,x) + \sum_ {n=1}^\infty \TPsi_n(\omega,x) \, ,
\end{equation}
with each term a Dyson series itself:
\begin{align} \label{eq:24a}
\TPsi_o(\omega,x) = \sum_{k=1}^\infty &\int_{-L}^{\infty} \! g_o(x,x_1) \cdots g_o(x_{k-1},x_k) V(x_1) \nonumber \\
& \, \,\, \cdots V(x_{k-1}) I(\omega,x_k) dx_1 \cdots dx_k 	\, ,
\end{align}
the series stemming from the first line of \eqref{eq:step2}, and the reflectivity terms, which can be re-arranged as,
\vspace{-9mm}
\begin{widetext}
\begin{align} \label{eq:25a}
\!\!\!\!\TPsi_n(\omega,x) = R^n \!(\omega) \! \sum_{k=n}^\infty  \! {1 \over n! (k-n)!}  \!\! \sum_{\sigma \in S_k}  \!\!\int_{-L}^{+ \infty} \! \!\!\!\! g_r(x_{\sigma(1)-1},x_{\sigma(1)}) \cdots    g_r(x_{\sigma(n)-1}&,x_{\sigma(n)})g_o(x_{\sigma(n+1)-1},x_{\sigma(n+1)}) \cdots  g_o(x_{\sigma(k)-1},x_{\sigma(k)})  \nonumber \\
 & \!\!\!\times V(x_1) \cdots V(x_{k-1})\, I(\omega,x_k)\, dx_1 \cdots dx_k ,
\end{align}
\end{widetext}

where $x_0 \! :=\! x$, $S_k$ is the permutation group of degree $k$ and ${1 \over n! (k-n)!}  \! \sum_{\sigma \in S_k} $ represents the sum on all possible distinct ways of ordering $n$ $g_r$'s and $k-n$ $g_o$'s, resulting in a total of ${|S_k| \over n! (k-n)!} = \binom{k}{n}$ terms. For instance, for $k=3$, $n=2$, we have
\begin{align}
& {1 \over 2! (3-2)!}  \nonumber \\
& \sum_{\sigma \in S_3} \! g_r(x_{\sigma(1)-1},x_{\sigma(1)}) g_r(x_{\sigma(2)-1},x_{\sigma(2)}) g_o(x_{\sigma(3)-1},x_{\sigma(3)}) \nonumber \\ 
&\,\,\,\,\,\,\,\,\,\,\,\,\,\,\,\,\,\,\,\,\,\,\,\,\,\,\,\,\,\,\,\,\,\,\,\,\,\,\,\,\,\,\, \,\,= \,\,g_r(x,x_1) g_r(x_1,x_2) g_o(x_2,x_3) \nonumber \\ 
&\,\,\,\,\,\,\,\,\,\,\,\,\,\,\,\,\,\,\,\,\,\,\,\,\,\,\,\,\,\,\,\,\,\,\,\,\,\,\,\,\,\,\,  \,\,\,\,+  g_r(x,x_1) g_o(x_1,x_2) g_r(x_2,x_3)  \nonumber \\
&\,\,\,\,\,\,\,\,\,\,\,\,\,\,\,\,\,\,\,\,\,\,\,\,\,\,\,\,\,\,\,\,\,\,\,\,\,\,\,\,\,\,\,  \,\,\, \,+  g_o(x,x_1) g_r(x_1,x_2) g_r(x_2,x_3) ,
\end{align}
which can be more eficiently obtained by mantaining the functions' arguments in the same position, but instead interchanging the relative positions of the $g_r\!$'s and the $g_o\!$'s.

\par
Without a doubt we've increased the mathematical complexity of the problem. Nonetheless, Eq. \eqref{eq:25a} has special significance: it's the frequency amplitude of the $n$-th \emph{echo} of the initial burst. There's no proper way to show this since there's no mathematical definition of an echo. However, with the following discussion and further application of this formalism to the Dirac delta potential, we hope to provide enough justification.

If $R = 0$ then $\TPsi = \TPsi_o$, the open system waveform, where only $g_o$ participates. Conversely, when we don't have a perfectly transmissible boundary ($R \neq 0$), we get an additional infinite number of Dyson series, as stated in Eq. \eqref{eq:tpsi}. These $\TPsi_n$ terms are expected to give a smaller contribution to $\TPsi$ as $n$ increases. This is mainly due to two features in \eqref{eq:25a}. 

\begin{itemize}

\item First, when $|R(\omega)| < 1$, $R^n\!(\omega)$ is obviously an attenuation factor with a greater impact at large $n$. It indicates $n$ partial reflections at the boundary, as physically done by the $n$-th echo. Moreover, echoes have the distinctive feature of being spaced by the same distance for any pair of successive echoes. The fact that $\TPsi_{\!(n+1)}$ has an additional factor of $R(\omega)$ than $\TPsi_{\!(n)}$, hence an, independent of $n$, phase difference of $\arg[R(\omega)]$, indicates this. 

\item Moreover, the fact that the Dyson series starts at $k = n$. Since $g_o$ and $g_r$ are of the same order of magnitude, it is natural to expect that the series starting ahead (with less terms) has a smaller magnitude and contributes less to $\TPsi$ than the ones preceding them. The additional term that $\TPsi_n$ possesses when compared to $\TPsi_{n+1}$, and can thus be used to evaluate their amplitude difference, is given by
\begin{align} \label{eq:diff}
\Delta_n&(\omega,x) = R^n\!(\omega) \! \int_{-L}^{ \infty} \!  \! g_r(x,x_1) \cdots  g_r(x_{n-1},x_n) \nonumber \\
& \,\,\,\,\,\,\,\,\,\, V(x_1) \cdots V(x_{n-1}) I(\omega,x_n) dx_1 \cdots dx_n \, .
\end{align}
\end{itemize}

Furthermore, latter echoes are seen to vibrate less than the first echoes. As stated before, since the Dyson series is basically an expansion on powers of $V/\omega^2$,  by starting at $k = n$, $\TPsi_n$ skips the high frequency contribution to the series until that point. This is intuitively due to high frequency signals trespassing the potential barrier more easily than lower frequency signals, which is the reason why high frequency behaviour predominates in the earlier echoes. 

\subsection{Inversion into the time domain}
Finally, let us make use of the inverse Laplace transform \eqref{eq:inverse} to obtain the time-dependent solution of wave equation \eqref{eq:1}. We start with the open system perturbation \eqref{eq:24a}. The frequency dependent terms are the Green's functions $g_o$, which have a pole at $\omega = 0$ (Eq. \eqref{eq:go}), and the source term $I$ which does not have any pole (Eq. \eqref{eq:I}). Thus, to keep the integrand convergent we should integrate above $\omega = 0$.  The frequency integral of the $k$-th term of \eqref{eq:24a} is
\be \label{eq:24c}
{1 \over 2 \pi i}\!\int^{+ \infty + i}_{- \infty + i} {e^{i \omega ( |x-x_1| + \dots + |x_{k-1}-x_k| - t)} \over  \omega^k } I(\omega, x_k)\, d \omega \, ,
\ee
where we've chosen $\beta = 1 \, (>0)$, in Eq. \eqref{eq:inverse}. The integrand is singular except when $k = 1$, due to the term $i \omega \psi_0(x)$ in $I(\omega,x)$, that cancels the $\omega$ in the denominator. For this term, we have
\be \label{eq:delta}
{1 \over 2 \pi} \int^{+ \infty + i}_{- \infty + i} e^{i \omega (|x-x_1|-t)} \, d \omega = \delta(|x-x_1| -t) \, ,
\ee
whereas for the term $- \dot{\psi}_0(x)$ of $I(\omega,x)$, we have to integrate a simple pole at $\omega = 0$,
\be \label{eq:theta}
- {1 \over 2 \pi i} \int^{+ \infty + i}_{- \infty + i} {e^{i \omega (|x-x_1|-t)} \over \omega} \, d \omega = \Theta(t-|x-x_1|) \, ,
\ee
which vanishes for $t < |x-x_1|$: The initial signal $\dot{\psi}_0(x_1)$ did not have enough time to travel to the point of observation $x$, i.e. these points are \emph{causally} disconnected.

Finally, integration of $\psi_0(x_1)$ and $\dot{\psi}_0(x_1)$ with \eqref{eq:delta} and \eqref{eq:theta}, respectively, yields the first term of $\Psi_o(t,x)$,  
\be \label{eq:psi0}
\Psi_i(t,x) = {1 \over 2} \bigg[ \psi_0(x-t) + \psi_0(x+t) + \int^{x+t}_{x-t} \! \dot{\psi}_0(x')\, dx' \bigg] \, .
\ee
If both $R(\omega)$ and $V(x)$ vanish, this is nothing but the final solution $\Psi(t,x)$. The equation above reveals that the initial waveform separates in two halves, propagating in opposite directions, just like an infinite plucked string.

For $k \neq 1$, we start by defining
\be
s_k := |x-x_1| + \dots + |x_k -x_{k+1}| - t \, ,
\ee
the causal distance, involving $k$ interaction points besides the point of observation $x$ and the source point $x_{k+1}$, for an elapsed time $t$.

With this definition the argument of the exponential in \eqref{eq:24c} is simply $i \omega s_{k-1}$. This integration, for $k \neq 1$, is
\be
{\Theta(-s_{k-1}) \over (k-1)!}\,{ \partial^{k-1} \over  \partial \omega^{k-1} } \Big[ e^{i \omega s_{k-1}} I(\omega,x_k)  \Big]_{\omega = 0} \, .
\ee
If $I(\omega,x_k)$ was independent of $\omega$, the term in brackets would be  $(i s_{k-1})^{k-1} I(x_k)$. But since $I$ has the linear form \eqref{eq:I}, we can write the term inside brackets as $(i s_{k-1})^{k-1} I(-i{ k-1 \over s_{k-1}},x_k)$ .

Putting everything together yields a Taylor-like expansion,
\begin{align} 
\label{eq:30}
\Psi_o(t,x) &= \Psi_i(t,x) -{1 \over 2} \sum_{k=1}^\infty {1 \over k!} \! \int_{-L}^\infty \Big({s_k \over 2}\Big)^k I\Big(\!-\!{i k \over s_k},x_{k+1}\Big) \nonumber \\
 & V(x_1) \cdots V(x_k) \, \Theta(-s_k) \, dx_1 \cdots dx_{k+1} \, .
\end{align}

Inversion of Eq. \eqref{eq:25a} follows the same lines. Instead of $s_k$, it is useful to define 
\begin{align}
s_{n,k} := &\, (x-x_1) + \dots + (x_{n-1}+x_n) \nonumber \\ 
& + |x_n - x_{n+1}| + \dots + |x_{k-1}-x_k| - t \, ,
\end{align}
in order to write the frequency integral, corresponding to inversion of the $k$-th term of \eqref{eq:25a} through \eqref{eq:inverse}, as
\be \label{eq:25c}
{1 \over 2 \pi i}\!\int^{+ \infty + i}_{- \infty + i}  {e^{i \omega s_{n,k}} \over  \omega^k } R^n\!(\omega) \, I(\omega, x_k)\, d \omega \, ,
\ee
where we replaced the Green's functions by their explicit forms \eqref{eq:go} and \eqref{eq:gr}.

Now, we can't go further unless we know $R(\omega)$ in detail. More specifically, its poles and divergent behaviour at $\pm i \infty$, which specify the choice of contour.

For completeness, we present below the calculation for $R$ given by Eq. \eqref{eq:Rb}: 
\red{
\begin{align} \label{eq:33}
&\Psi_n(t,x) = \delta_{n,1}\Psi_r(t,x) \nonumber \\
 &-{(-r)^n \over 2} \sum_{k=n}^\infty \! {1 \over n! (k-n)!}  \! \sum_{\sigma \in S_k}  \! \int_{-L}^\infty \!  {\big(\sigma(s_{n,k}) \!+\! 2 Ln\big)^{k-1} \over 2^{k-1}  (k-1)!}   \nonumber \\
 &\,\,\,\,\,\,\,\,\,\,\,\,\,\, V(x_1) \cdots V(x_{k-1})  I\Big(\!-\!{ i(k-1) \over \sigma(s_{n,k})\! +\!2 Ln},x_k\!\Big) \nonumber \\
 & \,\,\,\,\,\,\,\,\,\,\,\,\,\,\,\,\,\,\,\,\,\,\,\,\,\,\,\,\,\,\,\,\,\,\,\,\,\,\, \Theta\big(\!-\!\sigma(s_{n,k})\! -\!2 Ln\big) \, dx_1 \cdots dx_k,
\end{align}
with
\begin{align}
\sigma(&s_{n,k}) := \, (x_{\sigma(1)-1}-x_{\sigma(1)}) + \dots + (x_{\sigma(n)-1}+x_{\sigma(n)}) \nonumber \\ 
& + |x_{\sigma(n+1)-1} - x_{\sigma(n+1)}| + \dots + |x_{\sigma(k)-1}-x_{\sigma(k)}| - t \, ,
\end{align}
}
and
\be
\Psi_r(t,x) = - {r \over 2} \psi_0(t-x-2L)
\ee
the reflected initial waveform, present only in the first echo (due to the Kronecker delta $\delta_{n,1}$).
\par
The more critical reader may realize that this method is only useful if the explicit form of $R(\omega)$ is known. For instance, this is not the case for a wormhole system, where $R$ stands for the reflectivity of the Schwarzschild potential which can only be extracted numerically. Thus, one may ask if it is also possible to express the reflectivity of a generic potential as a perturbative series. The answer is yes.

\subsection{Reflectivity series}

If we send a wave from $+ \infty$  ($e^{-i \omega x}$), the reflectivity will be the factor of the reflected wave, $R(\omega) e^{i\omega x}$. The source term that corresponds to $\TPsi_0 = e^{-i \omega x}$ can be inspected from \eqref{eq:free}, with $g = g_o$ (we're trying to extract the reflectivity, so it's only natural to consider purely outgoing BCs at both sides), and is formally given by
\be
I(\omega,x) = 2 i \omega \lim_{l \to \infty}  \delta(x-l) \, e^{- i \omega l} \, \label{eq:Ir}
\ee
which non surprisingly corresponds to a source pulse located at $x \to \infty$. The factor of $e^{- i \omega l}$ takes care of the phase difference. 
\par
Now, with this source term, the solution, given by the Dyson series \eqref{eq:24a}, at $x \to \infty$ has the form $\TPsi(\omega,x \to \infty) = e^{- i \omega x} + R(\omega) \, e^{i \omega x}$, with
\begin{align} \label{eq:R2}
\!\!\!R(\omega) = \sum_{k = 1}^\infty {1 \over (2 i \omega)^k }&\int_{-\infty}^{\infty} e^{i \omega( - x_1 + |x_1-x_2|+ \dots + |x_{k-1}-x_k| - x_k)} \nonumber \\ 
 &\,\,\,\,\,\,\,\,\, \,\,\,V(x_1) \cdots V(x_k) \, dx_1 \cdots dx_k \, , 
\end{align}
the reflection coefficient expressed in terms of the potential, as promised.
\par
The above expression can also be used to compute the system's QNMs, which are the poles of $R(\omega)$. In fact, there's an ongoing discussion on whether the QNMs of the system with purely outgoing BCs at both sides (as in the above case) coincide with the ones where a mirror is introduced, thus replacing the outgoing BC at one side with other, more complex, BC. The mirror + potential system's QNMs should also be the poles of its "reflectivity". This concept, however, is not defined in the case both BCs are not purely outgoing, that is, if the system is only partially open. We can't simply take the initial wave as $\TPsi_0 = e^{- i \omega x}$, but instead, 
\be
\TPsi_0 = e^{- i \omega x} + R(\omega) e^{i \omega x} \, , \label{eq:tpsi0r}
\ee
where $R(\omega)$ is NOT the reflectivity of the system but the reflectivity associated with the non trivial BC at some $x = - L$. This is the correct form for $\TPsi_0$ since, by Eq. \eqref{eq:integral}, it should be the complete solution of the system when there's no potential barrier and also reduce to $e^{-i \omega x}$ when the mirror vanishes, that is, a free wave travelling to the left.  
\par
The reader may find comfort in this definition by noting that $\TPsi_0$ computed through Eq. \eqref{eq:free} with $I$ given by Eq. \eqref{eq:Ir} does indeed recover expression \eqref{eq:tpsi0r}.
\par
Naturally, the system's reflectivity is still defined by the factor multiplying the outgoing wave at $+ \infty$,  $\mathcal{R}(\omega) \, e^{i \omega x}$. With the source term \eqref{eq:Ir} and $g$ given by Eq. \eqref{eq:g}, we just need to evaluate expression \eqref{eq:dyson} at $x \to \infty$ to extract
\begin{align} \label{eq:R3}
&\mathcal{R}(\omega) = R(\omega) + {1 \over 2 i \omega} \sum_{k = 1}^\infty \int_{-L}^{\infty} (e^{- i \omega x_1} + R e^{i \omega x_1}) g(x_1,x_2) \cdots  \nonumber \\
&g(x_{k-1},x_k) (e^{- i \omega x_k} + R e^{i \omega x_k}) V(x_1) \cdots V(x_k) \, dx_1 \cdots dx_k \, ,
\end{align}
which reduces to $\mathcal{R} = R$ if the potential vanishes, and to Eq. \eqref{eq:R2} if $R \to 0$, as expected. We reinforce that the $R(\omega)$ in the above expression corresponds to the reflectivity associated with the non-trivial BC at $x = -L$ whereas, in \eqref{eq:R2}, it is the reflectivity of the potential barrier. We use the same letter for both since the mirror at $x = -L$ can be either due to a BC at this point, or a potential barrier, in this case computable through Eq. \eqref{eq:R2}.
\par
One can see that Eq. \eqref{eq:R3} does not diverge where Eq. \eqref{eq:R2} diverges, for arbitrary potential. In other words, the mirror+potential system and the completely open potential system do not share the same spectrum of QNMs. 
\par
In the next section, we'll apply this apparatus to a specific potential to explicitly see this difference.  

\section{The Dirac delta potential}

We now apply the previous formalism to the Dirac delta potential,
\be  \label{eq:V}
V(x) = 2 V_0 \, \delta(x) \, ,
\ee
with $V_0 > 0$. 

\subsection{Open system solution: $\Psi_o$}
Instead of employing Eq. \eqref{eq:30} straight ahead, it's interesting to first compute the frequency amplitude from Eq. \eqref{eq:24a}. The $k=1$ term corresponds to the freely propagating initial waveform, $\TPsi_i(\omega,x)$. For $k>1$, the $k-1$ delta functions collapse all the integrals except the integration in $x_k$,
\begin{align} \label{eq:36a}
\TPsi_o(\omega,&x) = \,\TPsi_i(\omega,x) \,+  \nonumber \\
&+ \sum_{k=2}^{\infty} \int_{-L}^{\infty} { e^{i \omega(|x|+|x_k|)} \over (2 i \omega)^k} (2 V_0)^{k-1} I(\omega,x_k)\, dx_k \, ,
\end{align}
which is in fact a $k$-independent integral: Relabeling $x_k \to x'$ and treating the sum as a geometric series,
simplifies the above to
\begin{align} \label{eq:tpsio}
\TPsi_o(\omega,x) =& \,\TPsi_i(\omega,x) \,+  \nonumber \\
&+ \int_{-L}^{\infty} { e^{i \omega(|x|+|x'|)} \over 2 i \omega} R_\delta(\omega) \, I(\omega,x')\, dx' \, ,
\end{align}
with
\be \label{eq:rdelta}
 R_\delta(\omega) = \sum_{k=1}^\infty \Big( {V_0 \over i \omega }\Big)^k = - {V_0 \over V_0 - i \omega} \, ,
\ee
the reflectivity of the Dirac delta potential \eqref{eq:V}, which could be directly computed from Eq \eqref{eq:R2}. It diverges at the QNM
\be \label{eq:qnm}
\omega = - i V_0 \, .
\ee
\par
With $\TPsi_o$ in hand we just have to apply Eq. \eqref{eq:inverse} to get the time-dependent solution:
\begin{equation} \label{eq:psiopendelta}
\Psi_o(t,x) = \Psi_i(t,x) - C_0(t\!-\!|x|)  + C_{V_0}(t \!- \!|x|) \, e^{-V_0(t-|x|)} \, ,
\end{equation}
with QNM excitation coefficient 
\begin{equation} \label{eq:C}
 C_{V_0}(t) = - {1 \over 2} \, \Theta(t) \int_{- t}^t e^{V_0 |x|} \, I(-i V_0,x) \, dx \, ,
\end{equation}
and $\Psi_i(t,x)$ given by \eqref{eq:psi0}.

Direct application of  Eq. \eqref{eq:30} would even be more straightforward: Instead of a geometric series, the infinite sum that factors out is the Taylor series of $e^{V_0(|x|+|x'|-t)}$.
~

Before we move on, we should point out the following. When there are no interactions, $V_0 = 0$, the two latter terms of \eqref{eq:psiopendelta} cancel each other and, as expected, $\Psi_o(t,x) = \Psi_i(t,x)$. Also note that, unlike conservative systems, the QNM excitation coefficient $ C_{V_0}$ is not a constant. One may then ask in what conditions does $\Psi_o(t,x)$ decay with the QNM behaviour for $t \to \infty$. We expect this to happen when $I(\omega,x)$ is sufficiently localized in space, corresponding to more "physical" sources. Even a decay $I(\omega,x) \sim e^{- a |x|}$, for some $a > 0$ gives $\Psi_o(t \to \infty, x) \sim e^{-a t}$. For a gaussian source $I(\omega,x) \sim e^{-a x^2}$ it is possible to rewrite the integrand in \eqref{eq:C} as $\sim e^{-a(x-b)^2}$ with $b = {V_0 \over 2 a}$. Even if the gaussian is disperse (small $a$), which makes $b$ assume large values, for $t \gg b$ the integrand will contribute little and $ C_{V_0}(t)$ is essentially independent of $t$. In the very limit $t \to \infty$, $ C_{V_0}$ will just be the integral of a gaussian in the real line, with convergent (and known) value and hence $\Psi_o(t \to \infty, x) =  C_{V_0} \, e^{- V_0 t}$.

\subsection{Echoes: $\Psi_n$}

To obtain $\Psi(t,x)$ we still need to get $\Psi_n(t,x)$, according to Eq. \eqref{eq:tpsi}. Since we haven't yet specified $R(\omega)$, we must start at the frequency amplitude and employ Eq. \eqref{eq:25a} with potential \eqref{eq:V}. 
\par
As in the previous case, the delta functions will collapse all integrals in the $k$-th term of expansion \eqref{eq:25a}, except the one in $x_k$, \red{implying that the Green's functions product in the integrand will have the form
\be
g_r(x,0)g_r(0,0) \cdots g_r(0,0) g_o(0,0) \cdots g_o(0,0) g_o(0,x_k),
\ee
for the identity permutation $\sigma(j) = j$. }
\par
Since $g_r(0,0) = g_o(0,0)$, according to Eqs. \eqref{eq:go} and \eqref{eq:gr}, a large number of permutations will turn out to be algebraically identical, more specifically, the ones involving interchanging the functions in the 'middle', with argument $(0,0)$. \red{Different terms arise when the pair of functions at the 'ends' is permutated. For instance, for the permutation $\sigma(n+1) = 1$ and $\sigma(n) = k$, instead we'd have
\be
g_o(x,0)g_r(0,0) \cdots g_r(0,0) g_o(0,0) \cdots g_o(0,0) g_r(0,x_k).
\ee
In fact, there are only 4 possible algebraically different outcomes for the pair of functions at both ends of the product, which make the sum ${1 \over n! (k-n)!}  \! \sum_{\sigma \in S_k}$ over the $g$'s simplify to}
\begin{align} \label{eq:43}
&g_r(x,0) g_r(0,x_k) \binom{k\!-\!2}{n\!-\!2} +  g_o(x,0) g_o(0,x_k) \binom{k\!-\!2}{n} \nonumber \\
&+ \big[ g_o(x,0) g_r(0,x_k) + g_r(x,0) g_o(0,x_k) \big] \binom{k\!-\!2}{n\!-\!1} .
\end{align}
To get the first term, for example, we have two $g_r\!$'s at the ends, leaving $k\!-\!2$ spots for the remaining $n\!-\!2$ $g_r\!$'s to be organized. The remaining terms follow the same reasoning. Albeit not directly apparent, we are dealing with a total of $\binom{k}{n}$ terms, as pointed out after Eq. \eqref{eq:25a}, since
\be
 \binom{k\!-\!2}{n\!-\!2} + 2\binom{k\!-\!2}{n\!-\!1} + \binom{k\!-\!2}{n} = \binom{k}{n}.
\ee
\par
Now, similarly to what happened in Eq. \eqref{eq:36a}, renaming $x_k \to x'$ will make the integral \eqref{eq:25a} independent of $k$ and a geometric-like series factors out for every term in Eq. \eqref{eq:43}. For $g_r(x,0)g_r(0,x')$, for instance, what factors out is
\be
\sum_{k= n}^\infty \binom{k\!-\!2}{n\!-\!2} \Big({V_0 \over i \omega}\Big)^{k-1} \!\!= \big[ R_\delta(\omega) \big]^{n-1}\!\!\!\! \!\!\!\!\!,
\ee
with $R_\delta$ given by Eq.\eqref{eq:rdelta}, where we've used the identity for the power of a geometric series,
\be
\sum_{k=n}^\infty  \binom{k\!-\!1}{n\!-\!1} \, r^k = \bigg[ \sum_{k=1}^\infty r^k \bigg]^{n}.
\ee
\par
Using the same identity for the remaining terms yields
\begin{align}
&\TPsi_n(\omega,x) = \int_{-L}^\infty \! \Big[ R_\delta^{n\!-\!1}\!(\omega) e^{i \omega(x+x')} + R_\delta^{n\!+\!1}\!(\omega) e^{i \omega(|x|+|x'|)}  \nonumber \\
&+  R_\delta^n \! (\omega) \big(e^{i \omega(x+|x'|)} + e^{i \omega(|x|+x')} \big)  \Big] R^{n}\!(\omega)  {I(\omega,x') \over 2 i \omega}  dx' .
\end{align}
\par
This variety of terms can be physically interpreted. The first one, with the product $R_\delta^{n\!-\!1} R^{n}$, corresponds to a wave sent left, towards the mirror, and also received from the mirror, travelling to the right. Take the second echo, $n=2$, for instance. It reflects first at the mirror, then at the delta, and again at the mirror, picking up a factor $R_\delta R^2$. 
\par
The second one, with $R_\delta^{n\!+\!1}R^n$, is the reverse situation. The wave is sent right, towards the delta, and then also received from the delta, but travelling to the left. This situation can only happen for $x \in [-L,0]$, when the observer is inside the cavity.
\par The same is true for the remaning terms, with $R_\delta^n  R^n$. Here, one of two situations happen. The wave is sent into the mirror and then received from the delta, or first sent into the delta and then received from the mirror, reflecting, in either case, an equal number of times at the delta and at the mirror.
\par
 The echoes' amplitude $\TPsi_n$ is similar, in form, to $\TPsi_0$. Besides the presence of $R^n$, the difference lies in the order of the pole of the QNM \eqref{eq:qnm}, due to the powers of $R_\delta$. Inversion will result in derivatives of the integrand, evaluated at the QNM. Thus, the echoes besides vibrating and decaying with the delta QNM, have a slightly different behaviour which we'll see below to be of the \emph{polynomial} sort. 
\par
To proceed with inversion, with the use of Eq. \eqref{eq:inverse}, we consider $R(\omega)$ given by Eq. \eqref{eq:Rb}, to get
\begin{widetext}
\begin{align} \label{eq:ecodelta}
\Psi_n(&t,x) = \delta_{n,1} {r \over 2}\bigg[ \!-\!\psi_0(t\!-\!x\!-\!2L) + \! \int_{-L}^{t-x-2L} \! \!I(0,x') dx' \bigg]  \nonumber \\
& -{(-r)^n \over 2} \Big( E_n(V_0;t-|x|-2Ln) + E_{n-1}(V_0;t-x-2Ln) -  E_n(0;t-|x|-2Ln) -  E_{n-1}(0;t-x-2Ln) \Big)
\end{align}
\end{widetext}
with
\begin{align} \label{eq:E}
&E_n(V_0;t)  =  \Theta(t)  {V_0^{n+1} \over n!} {\partial^n \over \partial V_0^n} \! \int_{-\min(t,L)}^t \! \!\!\!\! \! e^{V_0 (|x|-t) } {I(-i V_0,x) \over V_0} dx \nonumber \\
& +  \Theta(t\!+\!L) (1\!-\!\delta_{n,0}) {V_0^n \over (n-1)!} {\partial^{n-1} \over \partial V_0^{n-1}} \! \int_{-L}^t \! \!\!\! \!  e^{V_0 (x-t) } {I(-i V_0,x) \over V_0} dx  .
\end{align}
\par 
A few comments must be made. Interaction of the source with the delta potential is being accounted for in the first integral of Eq. \eqref{eq:E} whereas reflection at the mirror is being accounted in the second integral, hence the $\Theta$ functions ensuring that there is enough time for the source to reach the delta and the mirror, respectively. 
\par
The factor $(1 - \delta_{n,0})$ vanishes for $n=0$ and is $1$ otherwise. It's easy to see that it only vanishes for $\Psi_1$, the first echo, which instead possesses the term on the first line of Eq. \eqref{eq:ecodelta}, corresponding to the reflection at $x=-L$ of the left-travelling intial waveform. This is the only surviving term in case $V_0 \to 0$, when there is no cavity. 
\par
It is interesting to note that the integrals themselves do not depend on $n$, apart from the integration limits. The difference between echoes mostly lies in the order of the derivative on $V_0$. \red{For instance, when $\dot{\psi}_0 = 0$, the derivative will only act on the QNM exponential factor assigning the said \emph{polynomial} behaviour to the echoes' waveform:
\begin{align} \label{eq:E2}
&E_n(V_0;t)  =  \Theta(t)  {V_0^{n+1} \over n!} \! \int_{-\min(t,L)}^t \! \!\!\!\! \!\!\!\!\! \! (|x|-t)^n e^{V_0 (|x|-t) } \psi_0(x) dx \nonumber \\
& +  \Theta(t\!+\!L) (1\!-\!\delta_{n,0}) {V_0^n \over (n-1)!}  \! \int_{-L}^t \! \!\!\! \! (x-t)^n e^{V_0 (x-t) } \psi_0(x) dx  .
\end{align}
}
\par
Figure 3 shows a "time-lapse" of the complete waveform given by the sum of the open-system solution, Eq. \eqref{eq:psiopendelta}, with the first 3 echoes, described by Eq. \eqref{eq:ecodelta}, with initial condition
\be \label{eq:init}
\psi_0(x) = e^{-(x-10)^2} \,  , \, \, \, \, \dot{\psi}_0(x) = 0 \, .
\ee
and system parameters
\be \label{eq:parameters}
V_0 = 1 \, , \, \, \, \,  L = 10 \, , \, \, \, \,  r= 1 \, \, \text{(Dirichlet BC)}.
\ee

The complete sequence of events can be seen in video format at: \url{https://youtu.be/XfJNwuwbvnA} .

\subsection{QNMs}
To substantiate the discussion at the end of the first section, let us compute the full system's reflectivity. Usage of Eq. \eqref{eq:R3} with potential \eqref{eq:V} yields
\be
\mathcal{R}_\delta = R + (1+R) \sum_{k=1}^\infty \bigg[(1+R) {V_0 \over i \omega} \bigg]^k,
\ee
which simplifies to
\be
\mathcal{R}_\delta = {R_\delta + R + 2 R_\delta R \over 1 - R_\delta R}\, ,
\ee
by using the geometric series identity and definition of $R_\delta$, Eq. \eqref{eq:rdelta}.

It is easy to check that $\mathcal{R}_\delta \to R_\delta$ if $R \to 0$ and vice-versa, if $R_\delta \to 0$ then $\mathcal{R}_\delta \to R$. Moreover, if $R= - 1$, corresponding to a perfect mirror, then we should expect everything to be reflected back, independently of the potential. In this limit we can also see that $\mathcal{R}_\delta = -1$.

More interestingly, at the delta QNM \eqref{eq:qnm}, $R_\delta \to \infty$, the dependence on $R_\delta$ cancels to give $\mathcal{R}_\delta = -{1 + 2R \over R}$, which is finite for non-vanishing $R$. Thus, $\omega = - i V_0$ is not a QNM of the mirror+delta system. The QNMs are instead implicitly given by 
\be 
\label{eq:RR}
R(\omega_n) R_\delta(\omega_n) = 1 \, .
\ee
\begin{figure}[h] \label{fig:spectrum}
\includegraphics[scale=0.57]{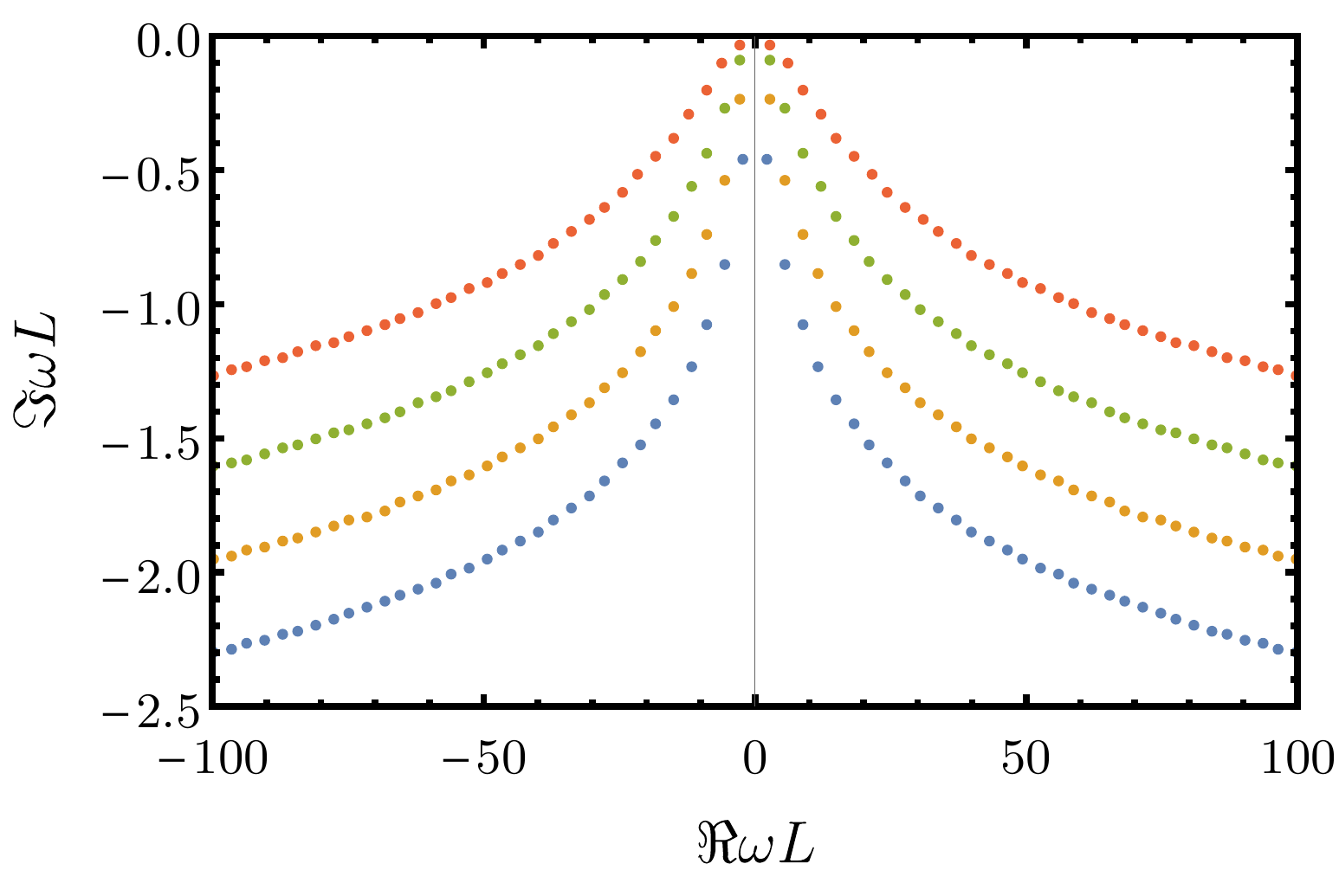}
\caption{QNM frequencies of the membrane-mirror system for different values of $V_0  L =1,2,4,8$ from bottom (blue) to top (red), respectively.}
\centering
\end{figure}
Fig. 1 plots the frequencies that respect the above, for $R$ given by Eq. \eqref{eq:Rb}, with $r = 1$ (Dirichlet BC at $x = -L$), which are well-approximated by the expression
\be
\omega_n = {n \pi \over L} - {1 \over 2 L} \arctan\!{n \pi \over L V_0} - 
{i \over 4 L} \log{\Big( 1+  { n^2 \pi^2 \over L^2 V_0^2}\Big)} \, .
\ee

Unsurprisingly, the imaginary part grows in magnitude with $|n|$. This implies that, at sufficiently long times, the perturbation will decay with the fundamental mode $\omega_{\pm1}$ (Fig. 2), even if the initial perturbation decays according to the pure-delta QNM~\eqref{eq:qnm} (as we show in Figure 3).
\begin{figure}[h] \label{fig:decay}
\includegraphics[scale=0.55]{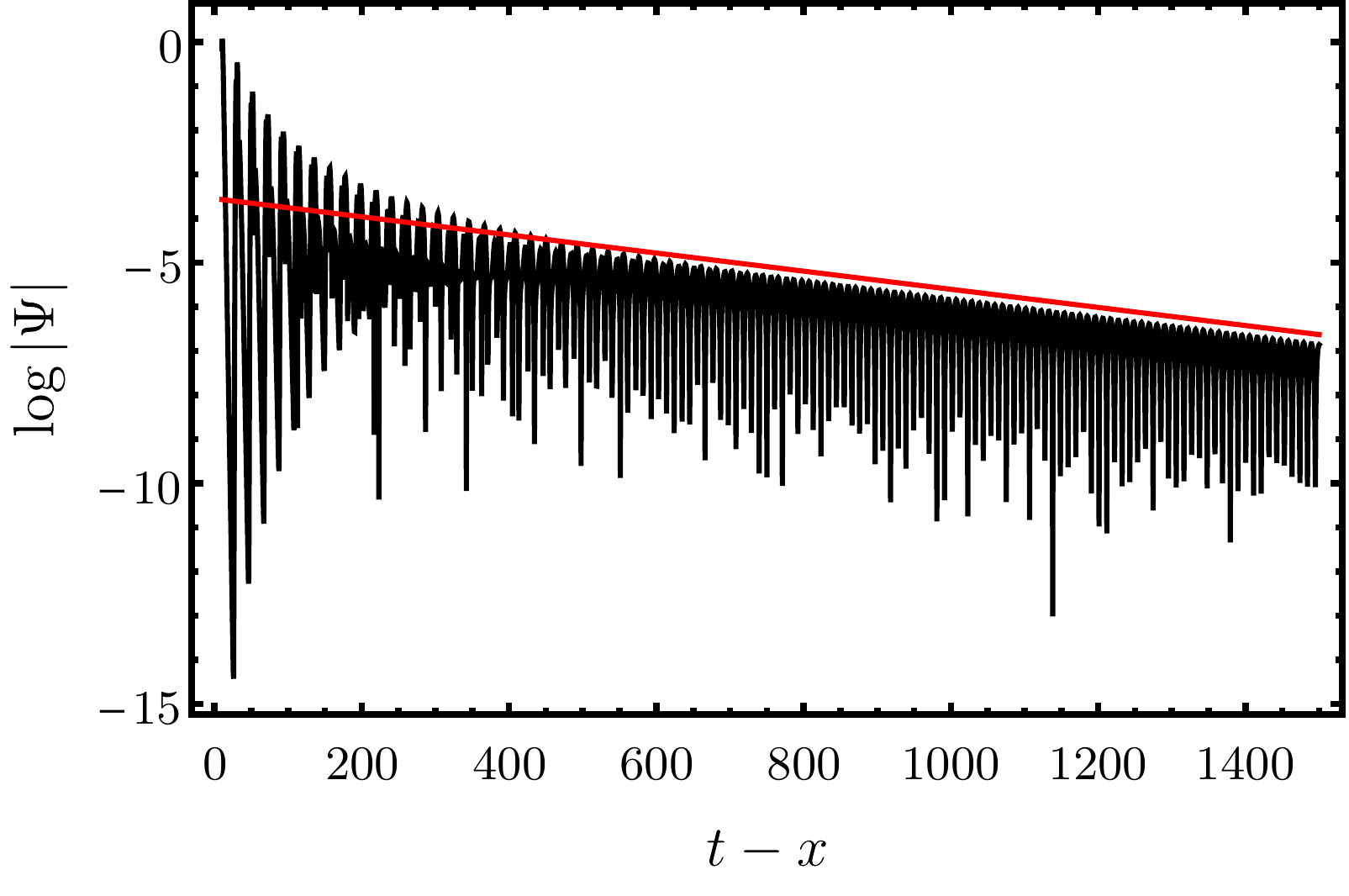}
\caption{Time evolution of the waveform using initial conditions \eqref{eq:init} and parameters \eqref{eq:parameters}. The plot shows the decay with the fundamental mode of the system (with mirror on the left), $\Im{\omega_{\pm1}} \approx - 0.00205$ (in red), for large $t$. The early echoes decay in a way that is governed by the QNMs of the {\it pure} delta. The high-frequency component is filtered out and progressively the signal is described by the modes of the composite system at late times, as it should.}
\centering
\end{figure}
We believe that this is the most convincing demonstration to date that the late-time decay is indeed governed by the QNMs of the composite system.

\begin{figure*}[ht]
\begin{tabular}{ccc}
\includegraphics[width=0.3\textwidth,clip]{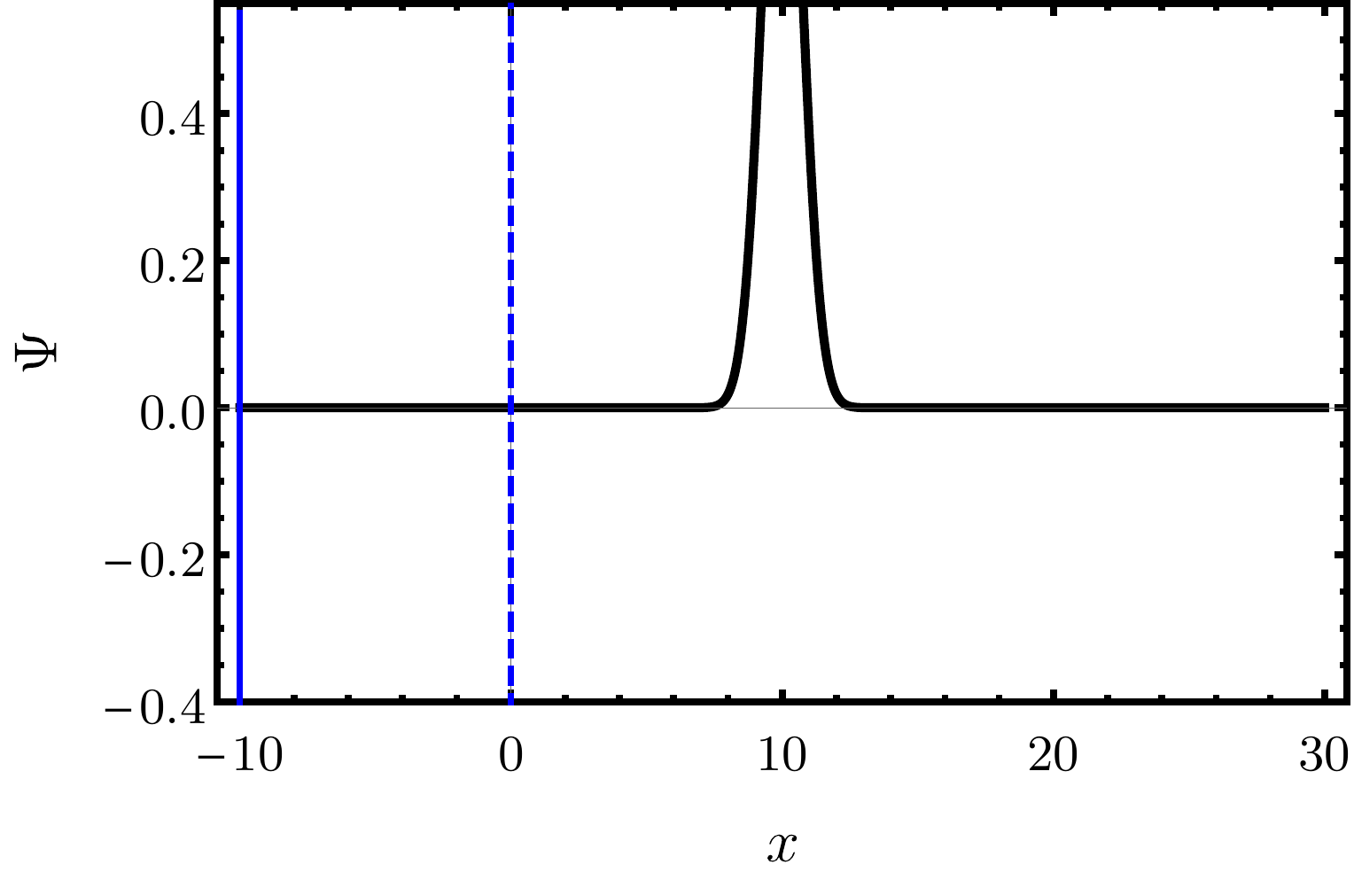}&
\includegraphics[width=0.3\textwidth,clip]{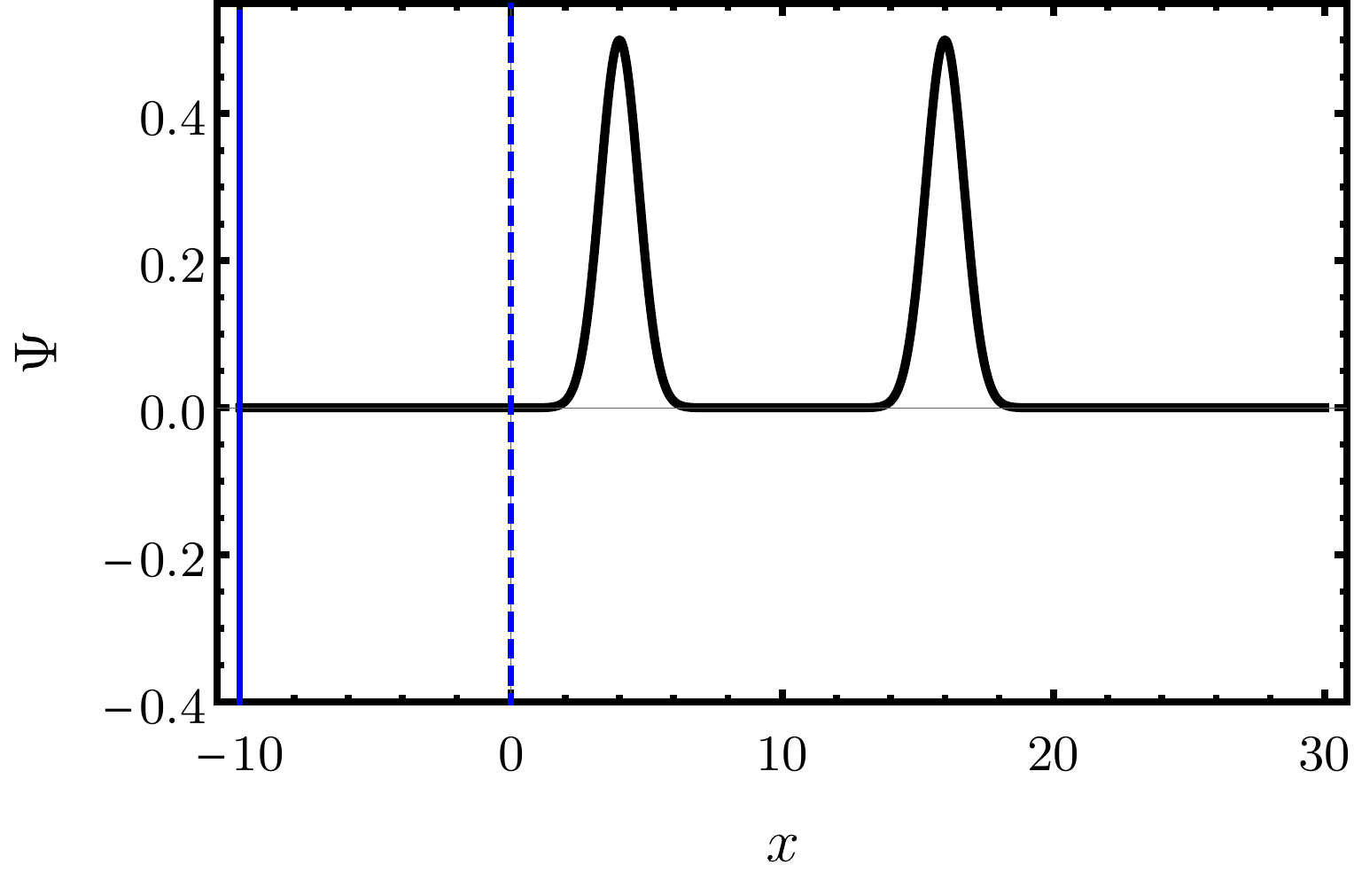}&
\includegraphics[width=0.3\textwidth,clip]{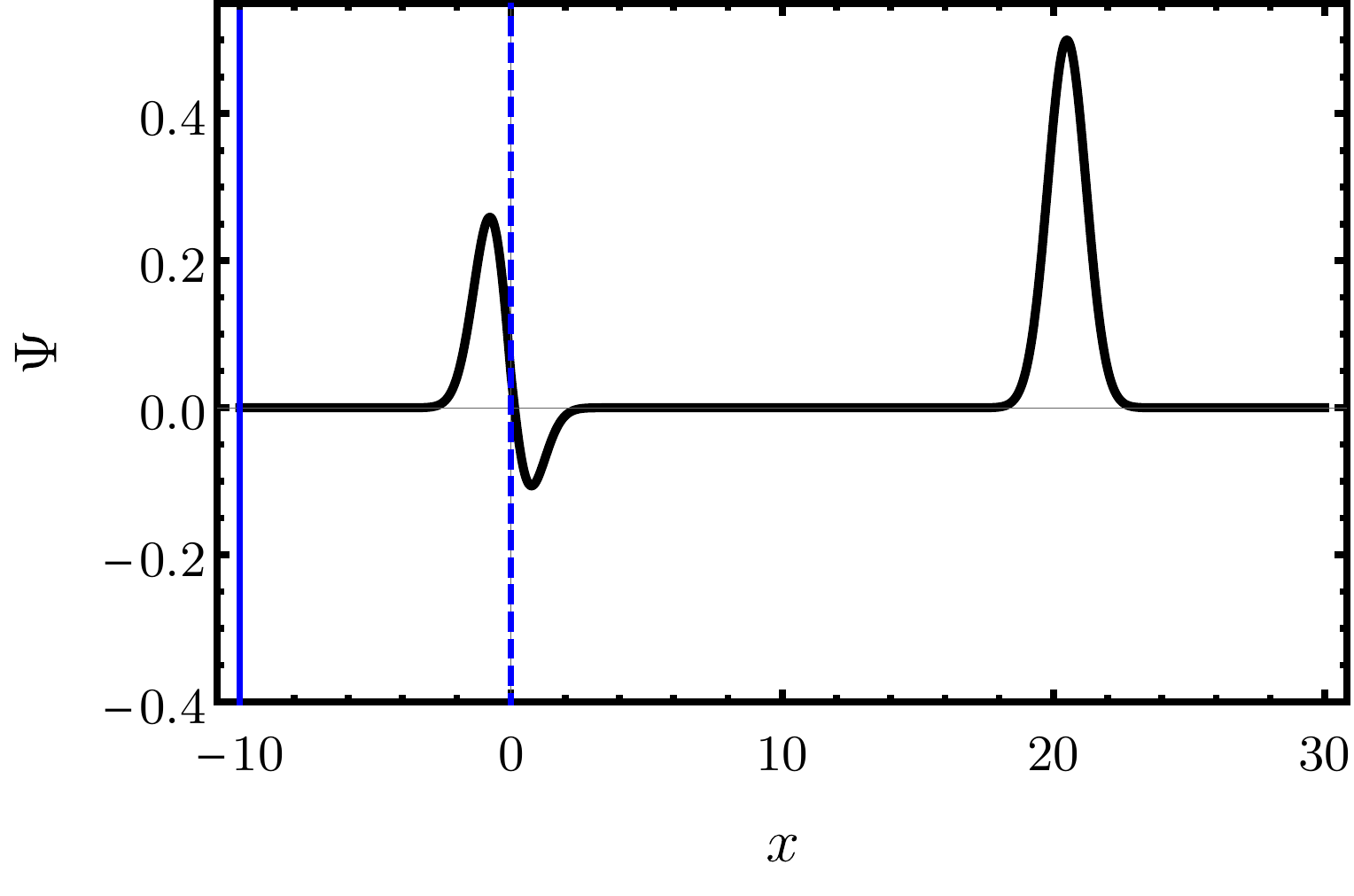}\\
\includegraphics[width=0.3\textwidth,clip]{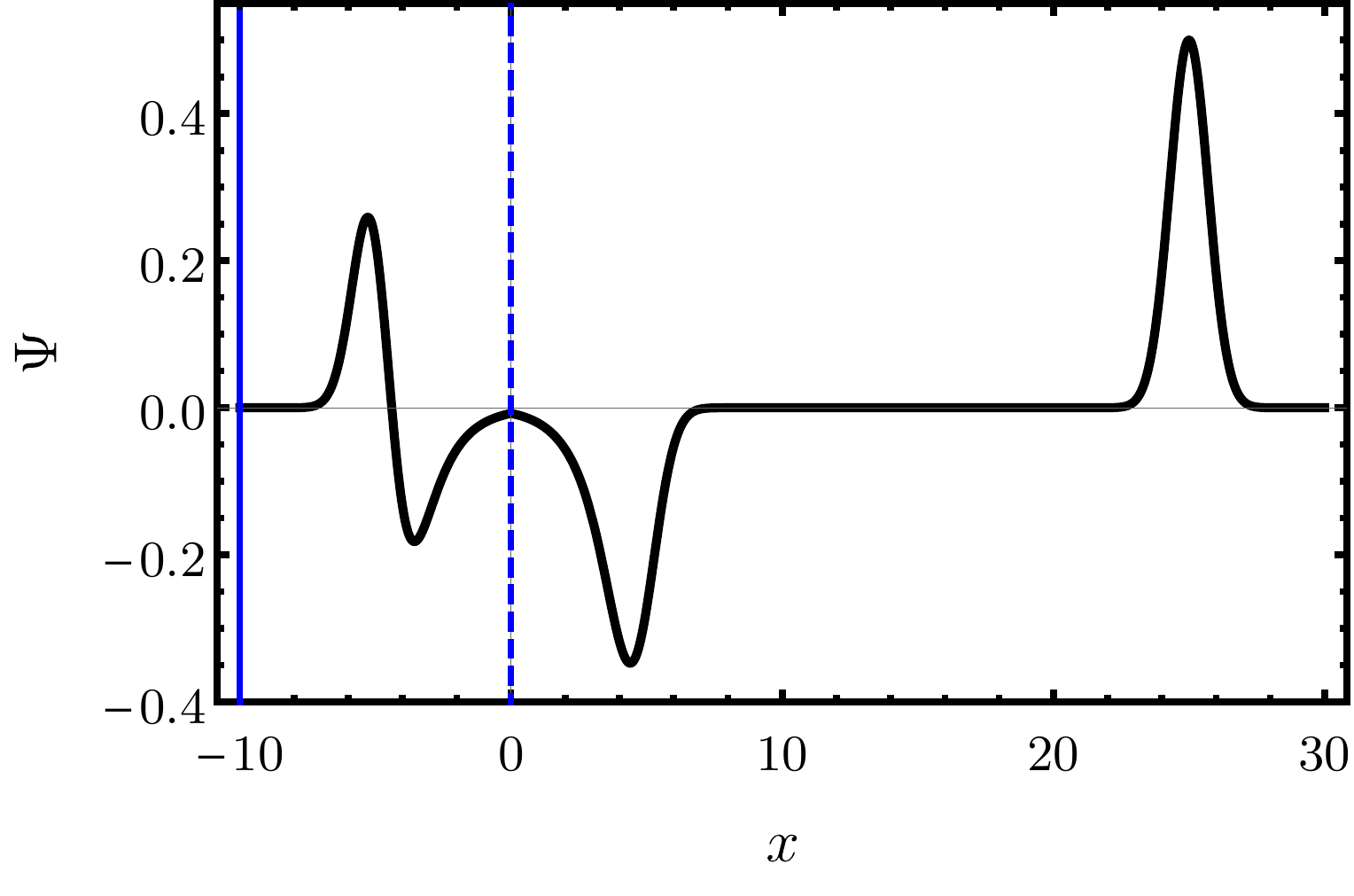}&
\includegraphics[width=0.3\textwidth,clip]{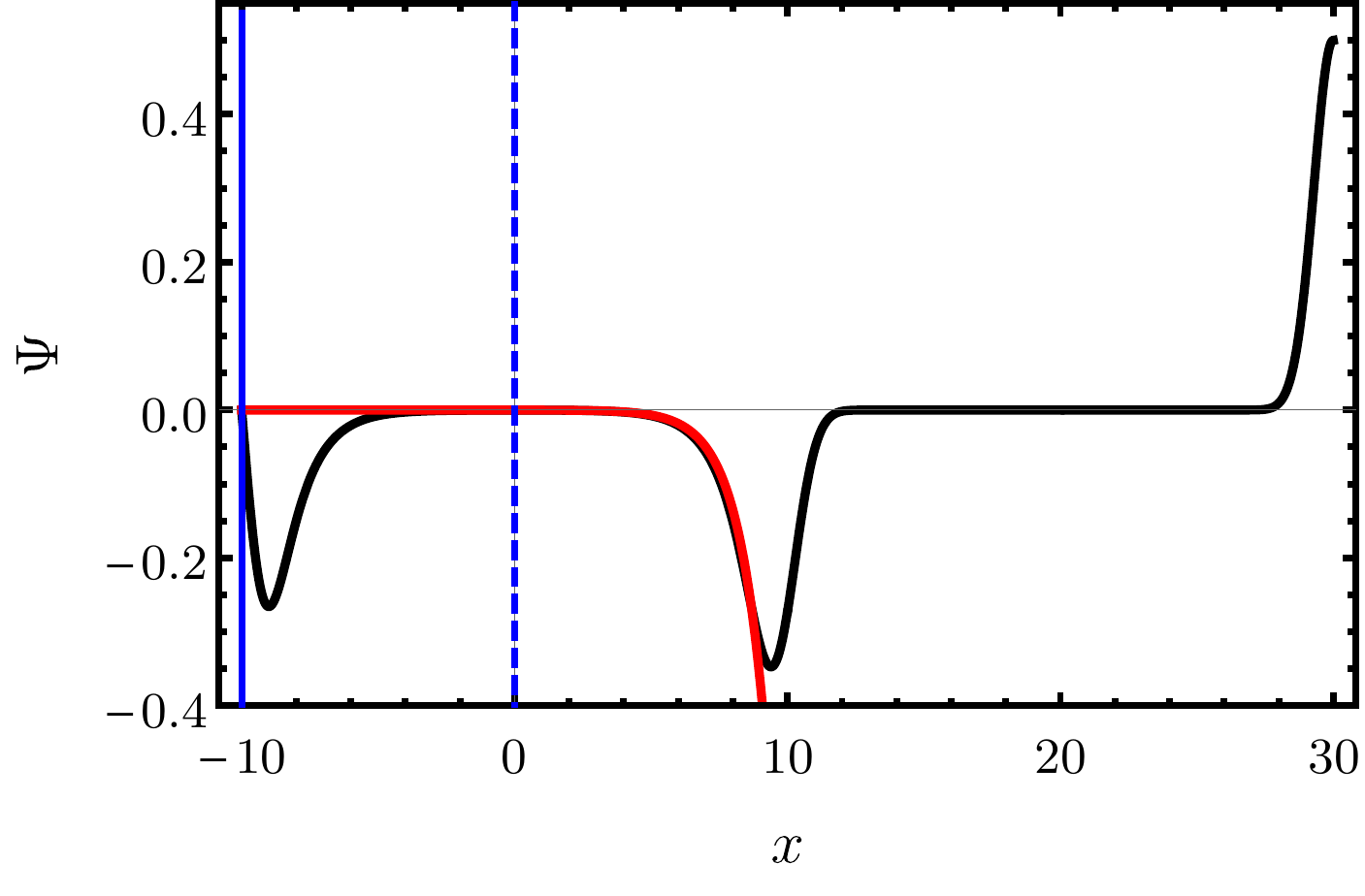}&
\includegraphics[width=0.3\textwidth,clip]{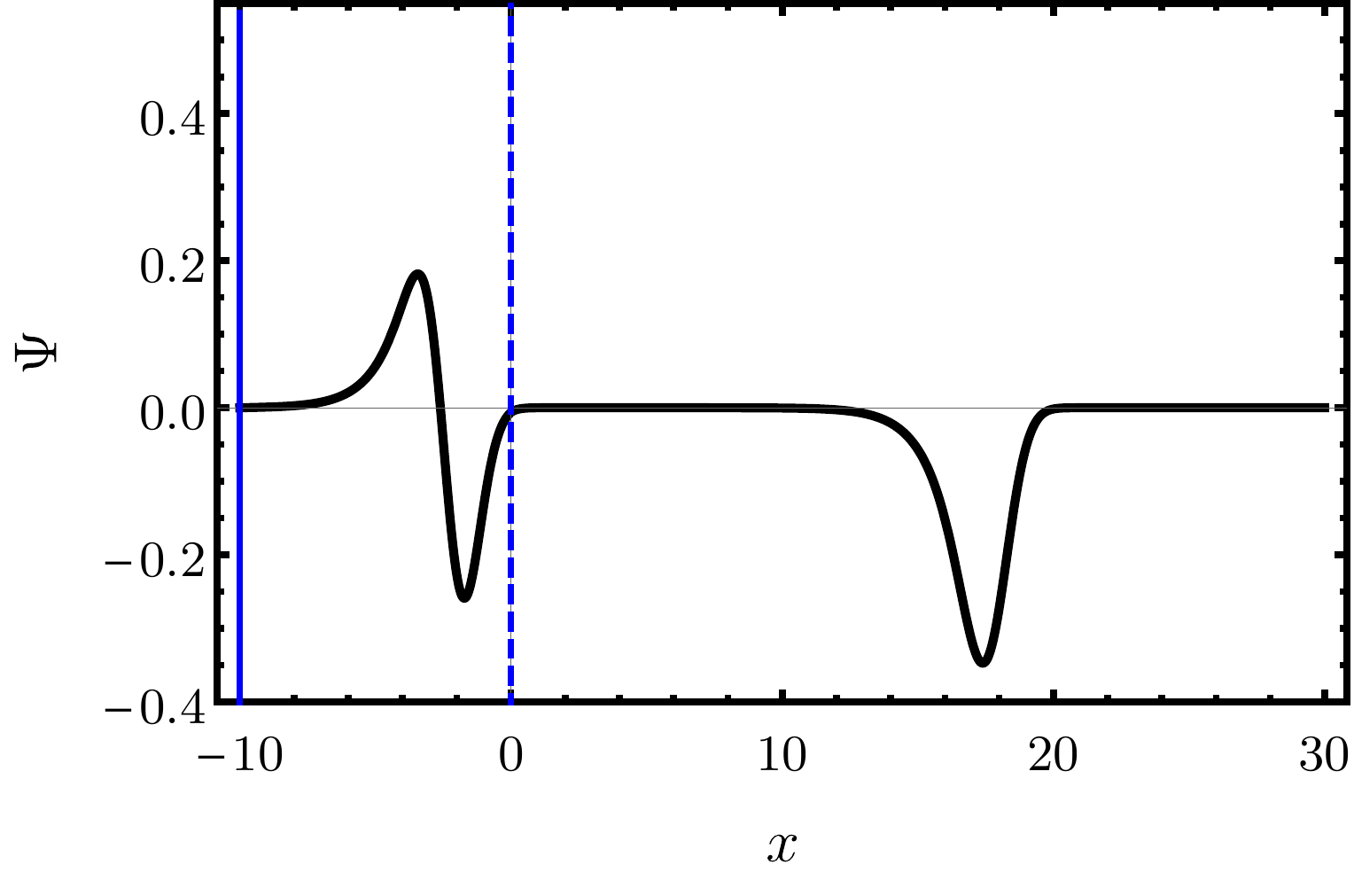}\\
\includegraphics[width=0.3\textwidth,clip]{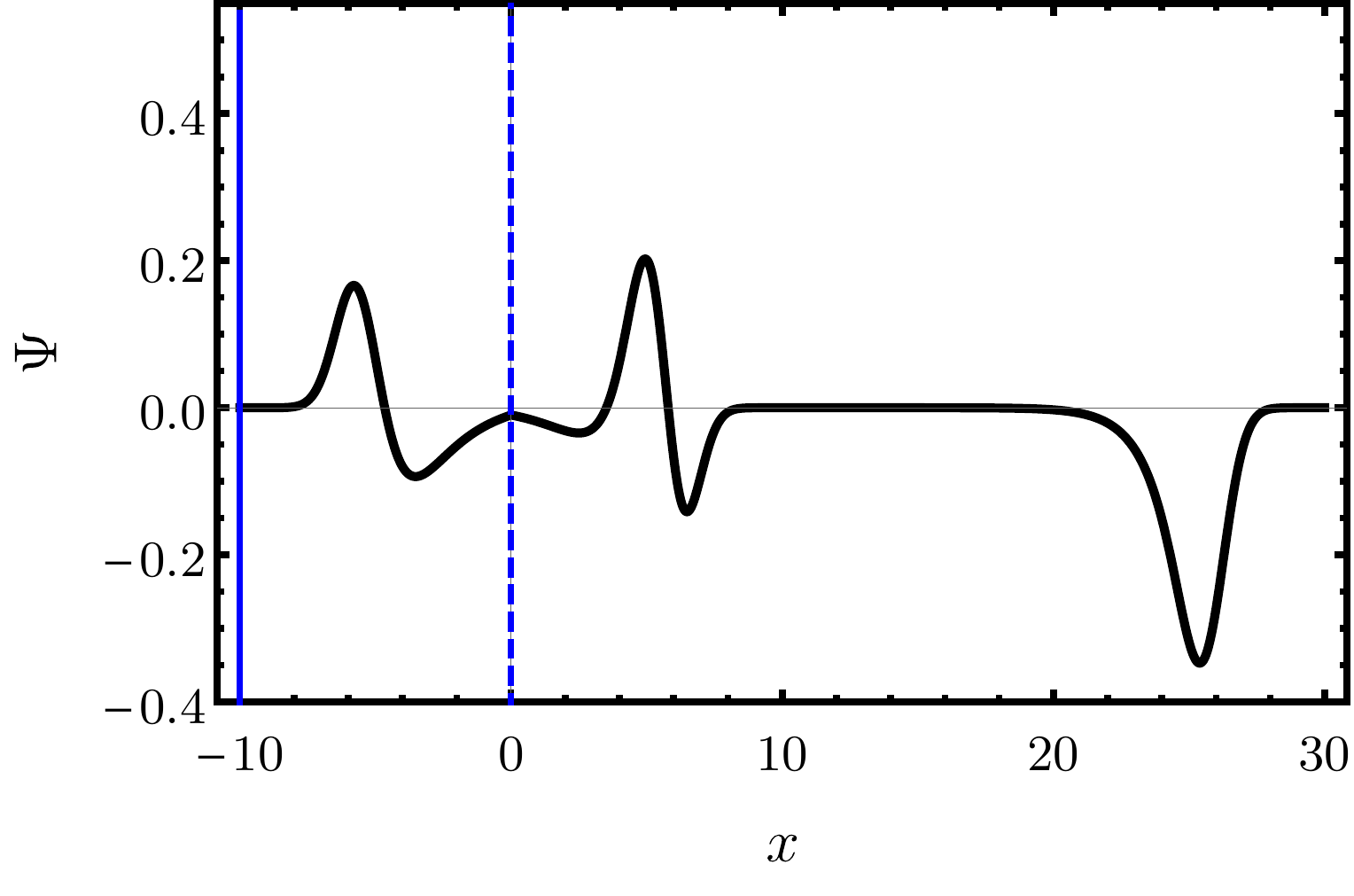}&
\includegraphics[width=0.3\textwidth,clip]{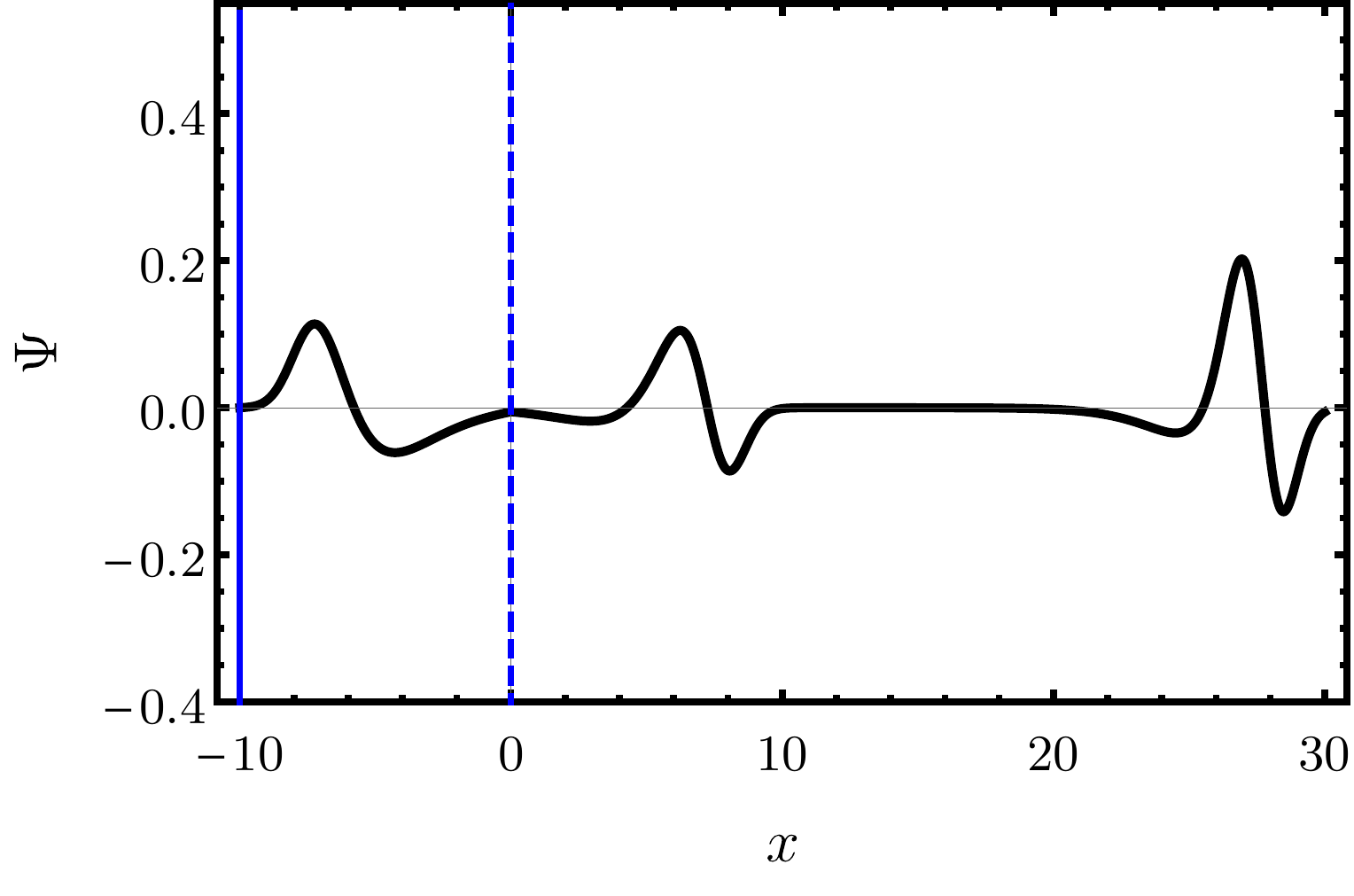}&
\includegraphics[width=0.3\textwidth,clip]{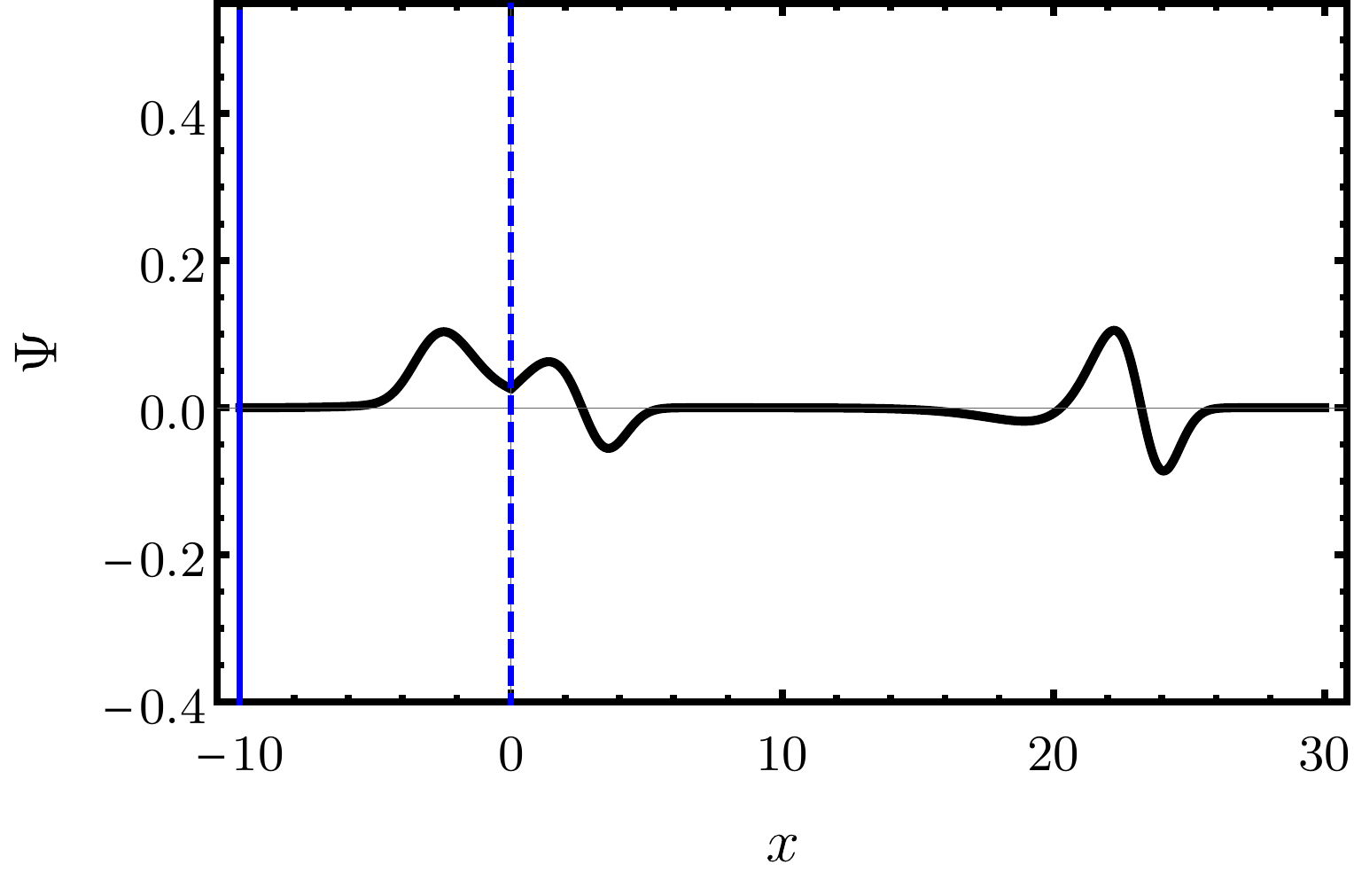}
\end{tabular}
\caption{Snapshots of the scalar profile at $t=0,6,10.5,16,20, 28,36,58, 74$ (top to bottom, left to right) in the presence of a delta-like potential at $x=0$ and a mirror at $x=-10$. The initial profile \eqref{eq:init} quickly gives way to two pulses traveling in opposite directions at $t=6$, as described by Eq.~\eqref{eq:psi0}; the left propagating pulse interacts with the (delta) potential at $t=10.5$ and gives rise to a transmitted pulse and a reflected one ($t=16$). The reflected pulse eventually reaches the boundary, at $t=20$, and will cross the potential at around $t=36$ giving rise to the first echo. The wave confined to the cavity (mirror+potential) will produce all subsequent echoes. At $t=58$, after $2L=20$ time units
the second echo emerges out of the cavity and at $t=74$ a third echo is about to be produced. These snapshots were obtained by adding three ``echoes,'' and coincides up to numerical error, with the waveform obtained via numerical evolution of the initial data.
In the central panel, the red line shows $- e^{V_0(x-10)}$, confirming that the initial decay is described by the QNMs~\eqref{eq:qnm} of the pure delta function (no mirror).}
\label{fig:echoes}
\end{figure*}
%
%
\section{Final remarks}
We have shown that a proper re-summation of the Dyson series solution of the Lippman-Schwinger equation
accounts for the presence of echoes in the waveforms of extremely compact objects (termed ``ClePhOs'' in the nomenclature of Refs.~\cite{Cardoso:2017njb,Cardoso:2017cqb}). We recover previous results, obtained with a completely different approach \cite{Mark:2017dnq}, but our approach also provides a few more insights. 
\par
\red{The key result Eq.\eqref{eq:25a} besides confirming the lower frequency content, decaying amplitude and constant distance of successive echoes, also explicitly relates the echoes waveform $\Psi_n$ with the initial conditions and sources incorporated into $I(\omega,x)$, the potential of the system $V(x)$, and the reflectivity of the wall $R(\omega)$ (the latter two function as the right and left sides of the \emph{lossy} cavity, respectively).  }
\par
\red{With the hypothetical future discovery of echoes in gravitational wave signals, the echo amplitude $\TPsi_n$ can be extracted up to experimental and numerical error. Together with the knowledge of $V(x)$ and $I(\omega,x)$, this turns Eq. \eqref{eq:25a} into an equation for $R(\omega)$, which encodes the information we currently lack on the quantum structure at the event horizon.}
\par
\red{Further developments of our formalism should include: application to other potentials besides the simple Dirac delta; a careful analysis of the convergence properties of Eq. \eqref{eq:25a}; extension of our methods to more than one spatial dimension; check if superradiant amplification is observed in Eq. \eqref{eq:25a} if $|R(\omega)| > 1$; implementation of the reflectivity series \eqref{eq:R2} and \eqref{eq:R3} to QNM computation; confirm the \emph{polynomial} behaviour of echoes in other systems besides the Dirac delta potential and use this information to echo modelling.}

\noindent{\bf{\em Acknowledgments.}}
%
We \red{thank Jos\'{e} Nat\'{a}rio for elucidating discussions and} are grateful to Kinki university in Osaka for kind hospitality while part of this work was being completed.
We acknowledge interesting discussions with the participants of the workshops ``Physics and Astronomy at the eXtreme (PAX)'' in Amsterdam and ``Gravitational Dynamics and Black Holes'' in Nagoya University.
V.C. acknowledges financial support provided under the European Union's H2020 ERC Consolidator Grant ``Matter and strong-field gravity: New frontiers in Einstein's theory'' grant agreement no. MaGRaTh--646597. Research at Perimeter Institute is supported by the Government of Canada through Industry Canada and by the Province of Ontario through the Ministry of Economic Development $\&$
Innovation.
This article is based upon work from COST Action CA16104 ``GWverse'', supported by COST (European Cooperation in Science and Technology).
This work was partially supported by FCT-Portugal through the project IF/00293/2013, by the H2020-MSCA-RISE-2015 Grant No. StronGrHEP-690904.

\bibliographystyle{apsrev4}
\bibliography{refs}

\end{document}